\documentclass[iop]{emulateapj}

\usepackage{graphicx}
\usepackage{epstopdf}
\usepackage{url}
\usepackage{amsmath}
\usepackage{rotating}
\usepackage{longtable}
\usepackage{multirow}
\usepackage{ctable}

\newcommand{\kepler}{\emph{Kepler}}

\slugcomment{Accepted by the Astrophysical Journal}

\shorttitle{Measuring Kepler's Detection Efficiency}
\shortauthors{Christiansen et al.}

\begin{document}


\title{Measuring Transit Signal Recovery in the Kepler Pipeline II: Detection Efficiency as Calculated in One Year of Data}

\author{Jessie L. Christiansen$^1$}
\author{Bruce D. Clarke$^2$}
\author{Christopher J. Burke$^2$}
\author{Shawn Seader$^2$}
\author{Jon M. Jenkins$^3$}
\author{Joseph D. Twicken$^2$}
\author{Jeffrey C. Smith$^2$}
\author{Natalie M. Batalha$^3$}
\author{Michael R. Haas$^3$}
\author{Susan E. Thompson$^2$}
\author{Jennifer R. Campbell$^4$}
\author{Anima Sabale$^4$}
\author{AKM Kamal Uddin$^4$}
\email{jessie.christiansen@caltech.edu}
\affil{$^1$NASA Exoplanet Science Institute, California Institute of Technology, M/S 100-22, 770 S. Wilson Ave, Pasadena, CA 91106}
\affil{$^2$SETI Institute/NASA Ames Research Center, Moffett Field, CA 94035}
\affil{$^3$NASA Ames Research Center, Moffett Field, CA 94035}
\affil{$^4$Wyle Laboratories/NASA Ames Research Center, Moffett Field, CA 94035, USA}




\begin{abstract}

The \kepler\ planet sample can only be used to reconstruct the underlying planet occurrence rate if the detection efficiency of the \kepler\ pipeline is known; here we present the results of a second experiment aimed at characterising this detection efficiency. We inject simulated transiting planet signals into the pixel data of $\sim$10,000 targets, spanning one year of observations, and process the pixels as normal. We compare the set of detections made by the pipeline with the expectation from the set of simulated planets, and construct a sensitivity curve of signal recovery as a function of the signal-to-noise of the simulated transit signal train. The sensitivity curve does not meet the hypothetical maximum detection efficiency, however it is not as pessimistic as some of the published estimates of the detection efficiency. For the FGK stars in our sample, the sensitivity curve is well fit by a gamma function with the coefficients $a=4.35$ and $b=1.05$. We also find that the pipeline algorithms recover the depths and periods of the injected signals with very high fidelity, especially for periods longer than 10 days. We perform a simplified occurrence rate calculation using the measured detection efficiency compared to previous assumptions of the detection efficiency found in the literature to demonstrate the systematic error introduced into the resulting occurrence rates. The discrepancies in the calculated occurrence rates may go some way towards reconciling some of the inconsistencies found in the literature.

\end{abstract}



\section{Introduction}
\label{sec:intro}

The \kepler\ Mission is a NASA Discovery Program mission designed to characterise the population of planetary systems using high precision photometric observations. The primary goal of the \kepler\ Mission is to measure $\eta_{\Earth}$, the frequency of Earth-size planets in the habitable zone of Sun-like stars. The spacecraft was launched in 2009, and for four years monitored the brightness of $\sim$192,000 stars nearly continuously, looking for the periodic dimmings indicative of transiting planets. The \kepler\ project has produced several planet candidate catalogues (Borucki et al. 2011a,b; Batalha et al. 2012; Burke et al. 2014; Rowe et al 2015; Mullally et al. 2015) from these data.

In order to determine the real, underlying population of planets from a sample of planet candidates, an essential step is to quantify the false negative rate: the fraction of the real planets that should have been detected that are not included in the sample (also called the survey completeness or survey detection efficiency). Initial analyses of the published \kepler\ planet candidate catalogues,  e.g. Borucki et al. (2011b; referred to as B11 for the remainder of this paper), \citet{Catanzarite2011}, \citet{Youdin2011}, \citet{Howard2012}, \citet{Dong2012} and \citet{Fressin2013} used various estimates of detection efficiency to constrain the occurrence rate of planets, but as yet there is no definitive empirical measure of this value for the \kepler\ pipeline. Petigura, Howard \& Marcy (2013; referred to as PHM13 for the remainder of this paper) used a custom-built pipeline to produce their own planet candidate sample from the \kepler\ data, which allowed them to directly quantify their detection efficiency and remove the uncertainties caused by estimating this quantity. We have initiated a large, robust study to empirically measure the false negative rate in the \kepler\ planet candidate sample, first introduced in \citet{Christiansen2013}, hereafter referred to as Paper I. In that study, we investigated the ability of the pipeline to preserve individual transit events, finding an extremely high fidelity of 99.7\% recovery of the expected signal strength across most of the investigated parameter space. This is extremely valuable for people performing their own transit searches using the \kepler\ light curves, as they can be assured that there has been little to no corruption of the signals at that point (barring the transits of very short period planets with periods below three days, or transits falling within two days of a gap in the \kepler\ data, as demonstrated in Paper I). 

The next step is to investigate the ability of the pipeline to recover periodic transit signals, as compared to individual transit signals. Here we present the results of our measurement of the detection efficiency of the \kepler\ pipeline across one year of long-cadence (30-minute integration) observations, comprising Q9--Q12 (for a detailed review of the \kepler\ Mission design, performance and data products, see Koch et al. 2010, Borucki et al. 2010 and Jenkins et al. 2010a)\nocite{Koch2010, Borucki2010, Jenkins2010a}. To measure the detection efficiency, we injected the signatures of simulated transiting planets into the calibrated pixels of $\sim$26,000 target stars across the focal plane, processed the pixels through the data reduction and planet search pipeline as normal, and examined the resulting detections. In Section \ref{sec:design}, we describe in detail the configuration of the tested pipeline and the generation and injection of the simulated planet signals. In Section \ref{sec:results} we examine the characteristics of the detections and generate the pipeline's sensitivity curve. In Section \ref{sec:discussion} we explore the impact of the sensitivity curve on the derived underlying planet population, in order to understand the systematic biases in the derived occurrence rates caused by assumptions about the pipeline sensitivity. Finally in Section \ref{sec:conclusions} we summarise the results and outline the further work required.


\section{Experiment Design}
\label{sec:design} 

The data reduction pipeline has been described in detail in a series of papers; for an overview see \citet{Jenkins2010b} and Figure 1 therein. The performance of the `front end' of the pipeline, comprising the modules Calibration (CAL: calibration of raw pixels; Quintana et al. 2010)\nocite{Quintana2010}, Photometric Analysis (PA: construction of the initial flux time series from the optimal aperture for each target; Twicken et al. 2010)\nocite{Twicken2010}, and Pre-Search Data Conditioning (PDC: removal of common systematic signals from the flux time series; Smith et al. 2012, Stumpe et al. 2012, Stumpe et al. 2014)\nocite{Smith2012,Stumpe2012,Stumpe2014}, in preserving transit signals was examined in Paper I. The performance of the full pipeline, including the `back end', comprising the modules Transiting Planet Search (TPS:  searching the light curves for periodic transit signals; Jenkins et al. 2010b, Seader et al. 2013)\nocite{Jenkins2010b,Seader2013} and Data Validation (DV: examination and validation of the resulting candidate signals against a suite of diagnostic tests; Wu et al. 2010)\nocite{Wu2010}, is examined here. In particular, the data products and software versions match those used to produce the Q1--Q16 Threshold Crossing Event (TCE) catalogue presented in \citet{Tenenbaum2013} and the associated Q1--Q16 Kepler Object of Interest (KOI) catalogue (Mullally et~al. 2015). Explicitly, the CAL and PA products were from Data Releases 12 (8.0), 13 (8.0), 15 (8.0) and 17 (8.1) for Quarters 9, 10, 11 and 12 respectively, where the SOC pipeline version is given in brackets after the release number; the PDC products were from Data Release 21 (8.3); and the TPS and DV products were produced using version 9.1 of the SOC pipeline. Using the more updated PDC and TPS/DV products in both the generation of the Q1--Q16 catalogue and this transit injection experiment allowed us to take advantage of the intervening updates (including Bayesian analysis to remove common systematic signals across target light curves in PDC and vetoing of non-astrophysical signals and iterative searching in TPS), without waiting for the long term re-processing of the CAL and PA products. Subsequently, updates to CAL and PA which have improved the detection efficiency in the meantime are not tested here; they will however be included in the Q1--Q17 catalogue and corresponding transit injection experiment. 

Some of the potential areas for signal loss in the pipeline are described in Paper I. One of these is signal distortion, by such processes as aperture errors and losses, and Sudden Pixel Sensitivity Drop-outs (SPSDs) occurring during transits. Additional causes of signal distortion include pipeline processes, such as the systematic error removal in PDC or the harmonic removal in TPS. Another source of signal loss is caused by signal masking: the pipeline now iteratively searches each flux time series down to the 7.1$\sigma$ signal detection threshold, removing the observations that contributed to the detected signals before searching again. This effectively reduces the number of cadences when transits can be observed, and as a consequence, the detectability of additional signals in that time series. 

In order to investigate the impacts of signal distortion and signal masking, we performed the following experiment. We selected 15 sky groups across the \kepler\ focal plane---a sky group refers to the set of target stars that fall together on the same CCD channel. As the \kepler\ spacecraft rotates around its boresight every three months, to keep the solar panels pointed at the sun, the set of stars rotates together onto a different CCD channel, and so on until after four rotations (one year of observations), the set of stars returns to the original CCD channel. The sky groups were selected to sample a range of the focal plane architecture and CCD channel characteristics in Q9 (in Q10--Q12, these sky groups will fall on other CCDs, typically with `average' behaviour); they are listed in Table \ref{tab:CCDs}. For computational reasons we limit this first multi-quarter transit injection experiment to one year of data comprising Q9--Q12. This is the first full year of spacecraft operations without a long data gap due to a spacecraft anomaly, and was chosen to isolate the impact of the data reduction pipeline on the detection efficiency; the impact of the window function due to long data gaps in the full data set will be investigated more thoroughly in the full baseline (Q1--Q17) transit injection experiment.

\begin{footnotesize}
\ctable[
  caption={{\small Sky groups and corresponding Quarter 9 CCD channels used in the transit injection test, and a qualitative description of any noted features of the channel. In the other quarters used in this test (Q10--Q12), the sky groups will fall on other CCDs, typically with `average' behaviour.}
  \label{tab:CCDs}}]{llllll}{
\tnote[a]{The measured bias in these channels has a much larger scatter from cadence to cadence than typical channels.}
\tnote[b]{A fixed pattern of varying-magnitude electronic crosstalk in the science pixels from clock crosstalk caused by the Fine Guidance Sensor (Kepler Instrument Handbook, 2009)}
\tnote[c]{A varying pattern (spatially and temporally) in the science pixels caused by a  temperature-dependent resonance in the Local Detector Electronics circuit (Kepler Instrument Handbook, 2009)}
\tnote[d]{A fixed pattern at the beginning of rows on some channels due to a voltage transient when parallel clocking out rows (Kepler Instrument Handbook, 2009)}
}{
\hline\hline
\small{Sky group} & \small{Channel} & \small{Description}\\
\hline
32 & 4 & Edge of field/worst focus\\
70 & 10 & Variable black/bias correction\tmark[a]\\
71 & 11 & Edge of field/worst focus\\
9 & 13 & FGS crosstalk\tmark[b]\\  
25 & 17 & Nominal/best focus\\
66 & 26 & Rolling band artefacts/Moir\'{e} pattern\tmark[c]\\ 
84 & 32 & Edge of field/worst focus\\ 
62 & 46 & Variable black/bias correction\\ 
78 & 50 & FGS crosstalk\\ 
4 & 56 & Edge of field/worst focus\\ 
18 & 58 & Rolling band artefacts/Moir\'{e} pattern\\ 
19 & 59 & Nominal/best focus\\
38 & 62 & Rolling band artefacts/Moir\'{e} pattern\\ 
74 & 70 & FGS crosstalk\\
53 & 81 & Start-of-line ringing\tmark[d]\\ 
\hline\hline}
\end{footnotesize}

Across these sky groups, we inject simulated transits into every target star. We generate our injected transits using the \citet{Mandel02} model. For each target star, the parameters of an initial estimated transit model are constructed from four observable parameters: (1) The signal-to-noise ratio (SNR) of a single transit is randomly drawn from a uniform distribution between 2$\sigma$ and 20$\sigma$; (2) The transit duration is randomly drawn from a uniform distribution between 1 and 16 hours (in the pipeline, we search for transit pulses with durations from 1.5--15 hours); (3) The impact parameter, $b$, is randomly drawn from a uniform distribution between 0--1; and (4) The phase of the first injected transit is randomly drawn from a uniform distribution between 0--1. The epoch of the first transit is also required to fall in Quarter 9, to ensure at least one injected transit per star. We then, using the calculations described in Appendix A of Paper I, generate a physical transit model which approximately reproduces those initial observed parameters, and from which the actual observed parameters are measured. We assume circular orbits and no limb darkening when generating the transit signal. We have included as electronic data both the physical and the measured observable parameters of the final injected planet models. Table \ref{tab:partablesample} shows an excerpt from the table to illustrate the contents.

\begin{sidewaystable}[h]
\footnotesize
\centering
\caption{Parameters of the injected transiting planets. The full table (10,341 rows) is available as an electronic data supplement to the paper.}
\begin{tabular}{lllllllllll}
\hline
\hline
\footnotesize{Kepler ID} & \footnotesize{Sky group} & \footnotesize{Period} & \footnotesize{R$_p$/R$_s$} & \footnotesize{a/R$_s$} & \footnotesize{$b$} & \footnotesize{$T_{\rm depth}$} & \footnotesize{$T_{\rm duration}$} & \footnotesize{Epoch}     &  \footnotesize{Expected MES} & Recovered\\
\footnotesize                     &  & \footnotesize{(days)}        &                                    &      &       & \footnotesize{(ppm)}              & \footnotesize{(hours)}                &\footnotesize{(BMJD)}&                                                & \\
\hline
 9755118 &   9 &  79.9530 & 0.0352 &  91.2298 & 0.5093 &   1241.473 &  5.206 & 55708.382 &   14.518 &  1 \\
 9755154 &   9 &   0.8870 & 0.0778 &   4.8476 & 0.2943 &   6047.453 &  1.087 & 55680.339 &  260.376 &  1 \\
 9755234 &   9 &  57.9840 & 0.0576 &  77.2773 & 0.2320 &   3322.075 &  4.457 & 55668.172 &   34.776 &  1 \\
 9815278 &   9 &   8.0340 & 0.0453 &  17.5103 & 0.1890 &   2055.826 &  2.725 & 55733.756 &  133.925 &  1 \\
 9815334 &   9 &   1.7550 & 0.0243 &   2.5050 & 0.5289 &    591.887 &  4.162 & 55723.432 &  164.277 &  1 \\
 9815427 &   9 &   1.8240 & 0.0246 &   4.8823 & 0.8932 &    606.471 &  2.219 & 55686.696 &  115.277 &  1 \\
 9815482 &   9 &  35.9650 & 0.0370 &  55.5882 & 0.9583 &   1370.606 &  3.843 & 55654.269 &   13.464 &  1 \\
 9815492 &   9 &  31.5930 & 0.0489 &  46.6409 & 0.2202 &   2390.012 &  4.024 & 55734.182 &   62.903 &  1 \\
 9815530 &   9 & 109.7020 & 0.0667 & 132.7673 & 0.2451 &   4454.328 &  4.908 & 55649.244 &   34.684 &  1 \\
 9815687 &   9 &   8.2130 & 0.0146 &  15.9514 & 0.1131 &    213.809 &  3.058 & 55731.764 &   19.592 &  1 \\
 9815837 &   9 &  50.5360 & 0.0360 &  62.9225 & 0.6047 &   1296.469 &  4.771 & 55649.738 &   27.238 &  1 \\
 9874912 &   9 &  18.9600 & 0.0272 &  34.5842 & 0.2569 &    738.354 &  3.256 & 55649.172 &   20.623 &  1 \\
 9875034 &   9 &  65.3840 & 0.0172 &  38.8429 & 0.8067 &    295.809 &  9.999 & 55642.542 &   11.171 &  1 \\
 9875070 &   9 &  28.5350 & 0.0452 &  52.5194 & 0.0158 &   2042.317 &  3.227 & 55685.497 &   70.409 &  1 \\
 9875085 &   9 & 103.4400 & 0.0140 &  81.0096 & 0.6069 &    196.574 &  7.585 & 55700.170 &    5.220 &  0 \\
 9875336 &   9 &   4.2350 & 0.0180 &  10.1619 & 0.4792 &    323.330 &  2.476 & 55704.299 &   20.937 &  1 \\
 9875410 &   9 &  44.4410 & 0.0163 &  30.1883 & 0.5857 &    265.188 &  8.745 & 55647.623 &   14.071 &  1 \\
 9875451 &   9 &  44.5600 & 0.0552 &  59.6182 & 0.1615 &   3044.727 &  4.440 & 55719.308 &   40.660 &  1 \\
 9875707 &   9 &   1.2590 & 0.0608 &   5.9426 & 0.6506 &   3693.535 &  1.258 & 55724.124 &  238.038 &  1 \\
 9875793 &   9 & 118.8520 & 0.0270 & 119.2722 & 0.3738 &    728.677 &  5.919 & 55651.631 &   19.200 &  1 \\
 ... & & & & & & & & & \\
\hline
\hline
\end{tabular}
\label{tab:partablesample}
\end{sidewaystable}

We inject the generated transit model into the calibrated pixels for the target in question. These modified pixels are then processed through the pipeline as normal. As in Paper I, the only departure from standard operations is that the motion polynomials (used for calculating the location of the target) and the cotrending basis vectors (used in the correction of systematic errors) are generated from a `clean' pipeline run. This is to avoid corruption from the presence of the injected transits, since the motion polynomials and cotrending basis vectors are generated from the data themselves, and will be distorted by the presence of transit signals in every single target.

In summary, the final order of processing is that we ran the original calibrated pixels (the output of CAL) of Q9--Q12 through PA, PDC, and TPS, without any modification, to generate the motion polynomials, the cotrending basis vectors, and the root-mean-square Combined Differential Photometric Precision (CDPP) for each target. We then injected the simulated transit signals into the calibrated pixels, one planet for every target across the 15 sky groups, and re-ran the modified pixels through PA, PDC, TPS and DV, utilizing the previously generated information as described.

\section{Results}
\label{sec:results}

\subsection{Detection efficiency}

\begin{figure}[t!]
\centering
\includegraphics[width=\columnwidth]{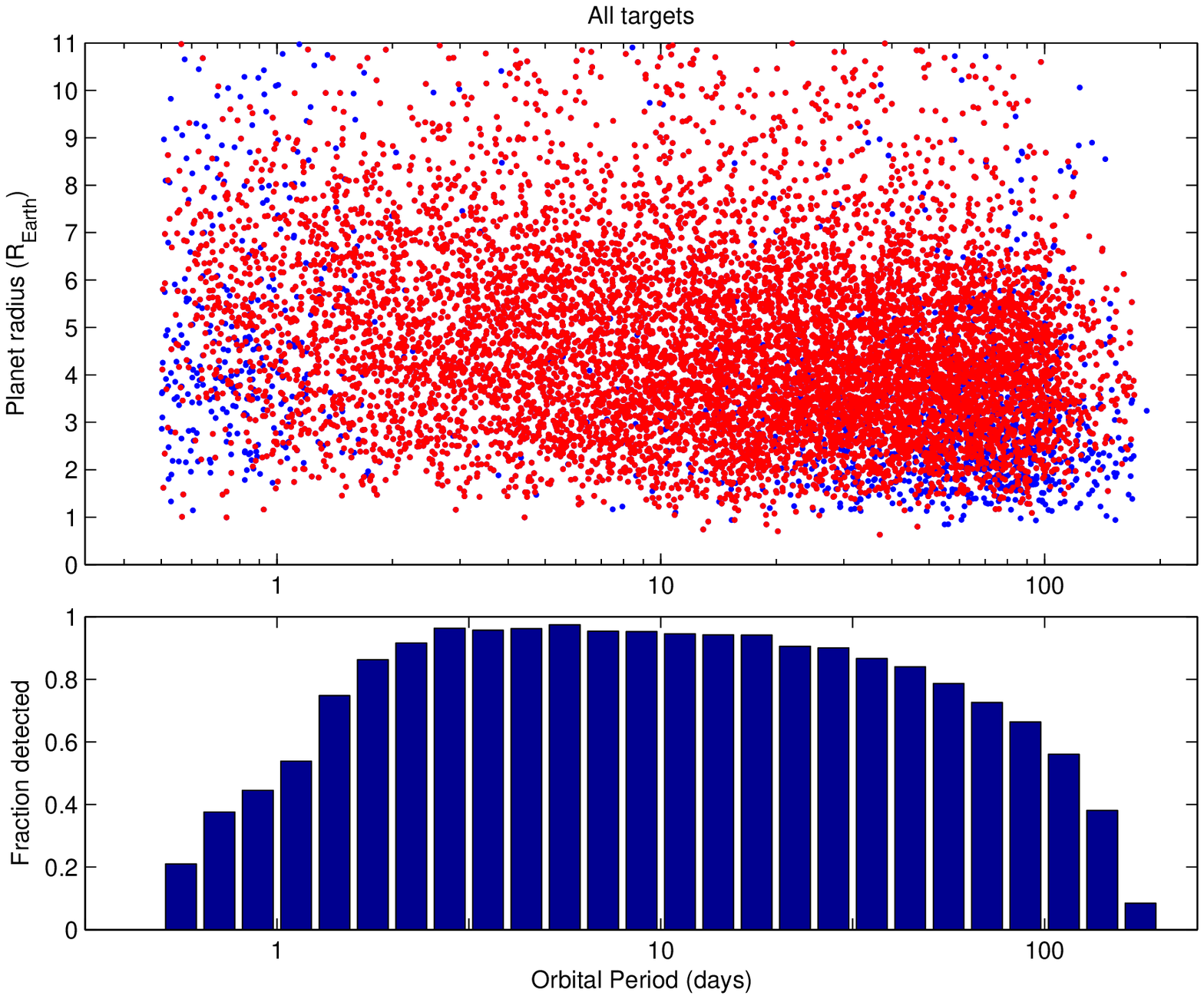}
\includegraphics[width=\columnwidth]{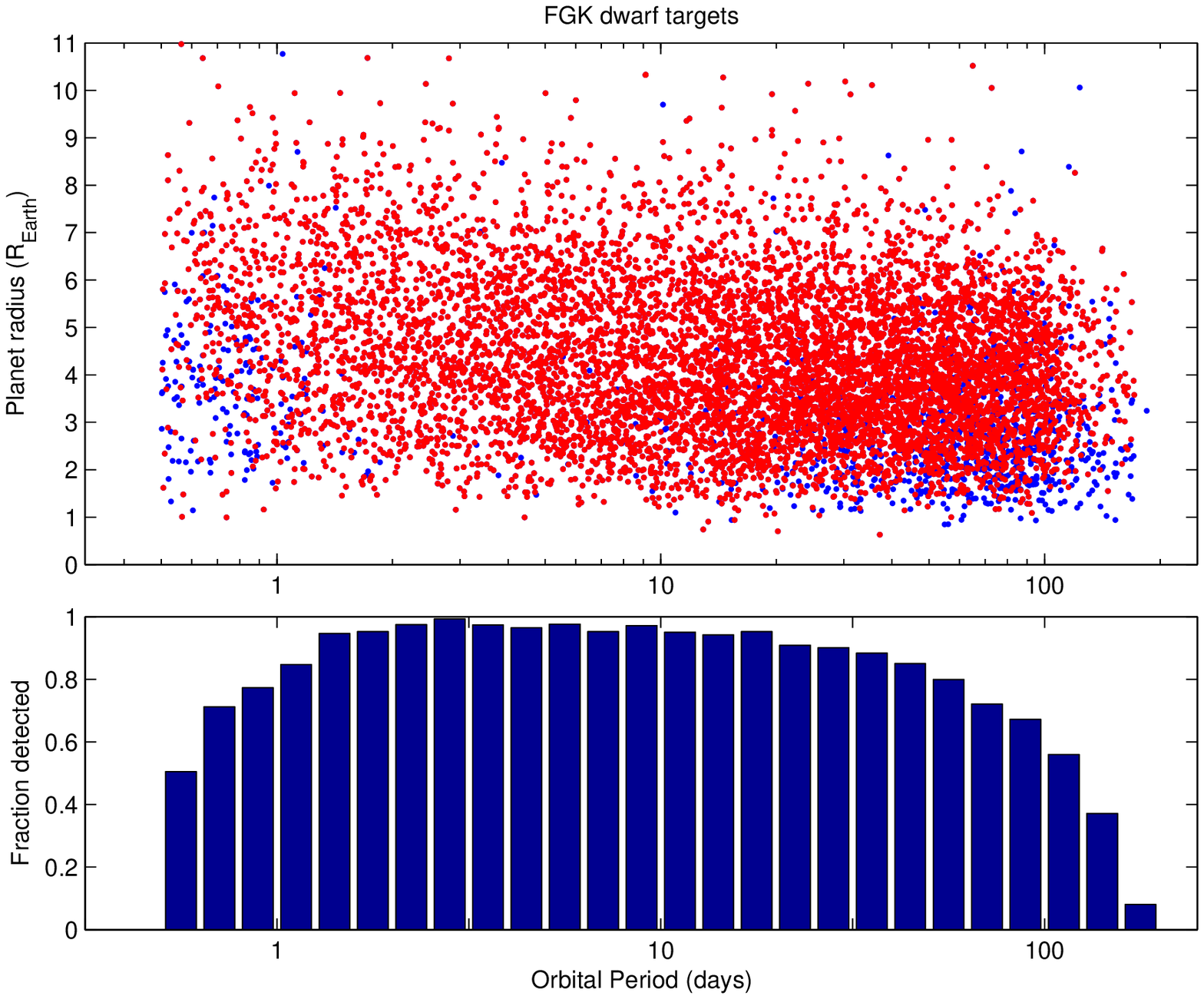}\caption{The distribution of parameters of the injected and recovered transit signals. {\it Upper two panels:} For all targets; {\it Lower two panels:} For the FGK targets. In both scatter plots, the blue points show the signals that were not successfully recovered, and the red points show the recovered signals.  The histograms show the fraction of recovered signals as a function of period.}
\label{fig:injectedparameters}
\end{figure}

\begin{figure}[h!]
\centering
\includegraphics[width=\columnwidth]{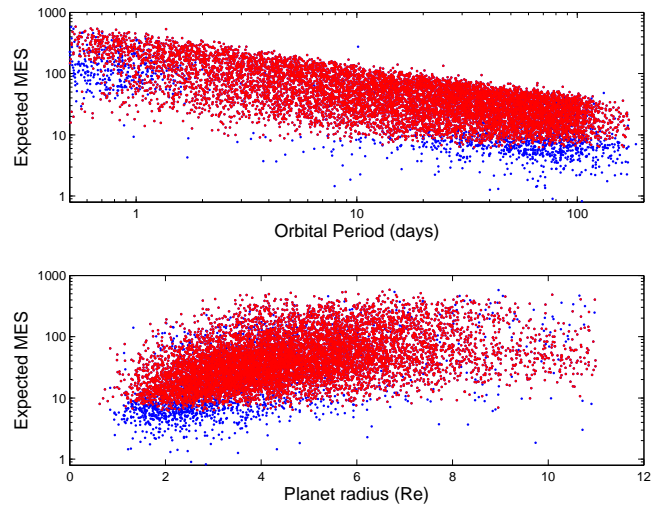}
\caption{The signal strength, measured in expected MES, of the injected transiting planets. In both panels, the blue points show the signals that were not successfully recovered, and the red points show the recovered signals. {\it Upper}: The correlation with orbital period, following the expected $\sqrt{N}$ dependence, where $N$ is the number of transits; {\it Lower}: The correlation with planet radius.}
\label{fig:mesdependence}
\end{figure}

The top panel of Figure \ref{fig:injectedparameters} shows the distribution of injected planet parameters for all targets across the planet radius range 0--11$R_{\oplus}$ and the planet orbital period range 0.5--200 days; there are 10,341 planet injections in this parameter space, of which 9,123 are successfully recovered by the \kepler\ pipeline (shown in red), and 1,218 are not (shown in blue). The histogram below shows the fraction of the injected signals that were successfully recovered as a function of period. Note that some of the injected signals are not expected to be successfully recovered given their expected SNR or orbital period, but are included as we wish to probe the sensitivity of the transition region between detection and non-detection. A `successful' recovery was defined as a signal being identified with an epoch within 0.5 days of the injected planet epoch, and a period within 3\% of the injected planet orbital period. Given the baseline of the observations (372 days) and the requirement of three transits for detection in the pipeline, the maximum detectable period is $\sim$185 days. The relatively small number of planets injected with radii $<$2$R_{\oplus}$ is due to the limiting of the SNR of individual transits to be $>$2$\sigma$. Figure \ref{fig:mesdependence} shows the signal strength of the injected planets over the orbital period (upper panel) and planet radius (lower panel) ranges of the injections, measured as the expected Multiple Event Statistic (MES) of the signal. The MES gives the significance of the correlation between the data and a putative box-shaped transit signal of a given orbital period, transit duration and phase in units of the uncertainty in the data \citep{Jenkins2010b}. In the upper panel, a linear fit to the data is overlaid, with a slope of -0.49, demonstrating the expected dependence of the signal strength on $\sqrt{N}$, where $N$ is the number of transits and is inversely proportional to the orbital period. 

There are two dominant effects visible in Figure \ref{fig:injectedparameters}: the first is the drop-off in detectability at very short (predominately $<$3-day) periods; the second effect is the drop-off in detectability at smaller radii and longer periods, which is the signal-to-noise detection threshold that we want to examine further.  

The cause of the drop-off in detectability at very short periods, in all examined cases, was the behaviour of the harmonic fitter in TPS, as described in Paper I. Before the flux time series is whitened and searched for periodic signals, a sinusoidal harmonic filter is applied to remove periodic stellar activity, allowing the pipeline to search variable stars for transit signals \citep{Tenenbaum2012}. Given the artificial separation of the injected transits in Paper I, we were limited in our ability to characterise the period parameter space where the harmonic fitter removes signal. With this extended analysis, we find that the impact of the harmonic fitter results in the non-detection of $\sim$60\% of the signals we would expect to detect between 0.5--0.6 days, and $\sim$40\% between 0.6--1.0 days, and drops  from there to 10\% at 1.5 days and 1\% at 2.0 days. 

We now turn our attention towards the drop-off in detectability with decreasing signal-to-noise, in order to measure the true detection threshold of the \kepler\ pipeline. In the bulk of our analysis, we limit the stellar sample to FGK stars (4000K $< T_{\rm eff} <$ 7000K, log $g>4.0$). This is the sample for which the project is currently calculating the occurrence rate of planets; variability in more active or more evolved stars can impact detectability, as shown below. There are 8,579 transit injections around the FGK sample across the planet radius range 0--11$R_{\oplus}$ and the planet orbital period range 0.5--200 days. The bottom two panels of Figure \ref{fig:injectedparameters} shows the distribution of injected and successfully recovered planet parameters for the FGK stars; 7,696 of the 8,579 injections were recovered. The histogram shows the fraction of the injected signals that were successfully recovered as a function of period; note here that the attenuation in detectability at short periods is mitigated somewhat. The attenuation in the full sample is somewhat driven by the more variable photometry of the giant stars on these short period timescales.

\begin{figure}[h!]
\centering
\includegraphics[width=\columnwidth]{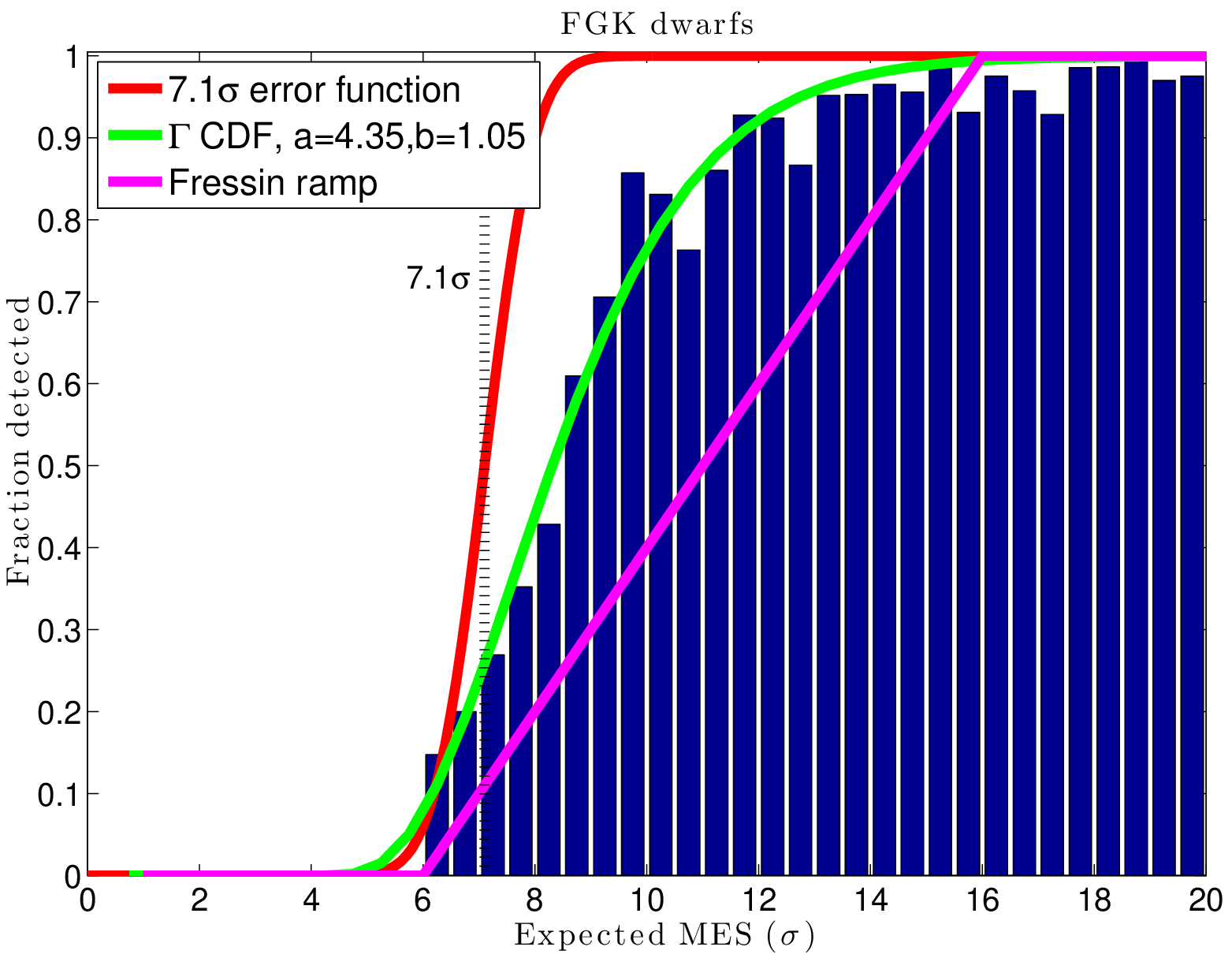}
\includegraphics[width=\columnwidth]{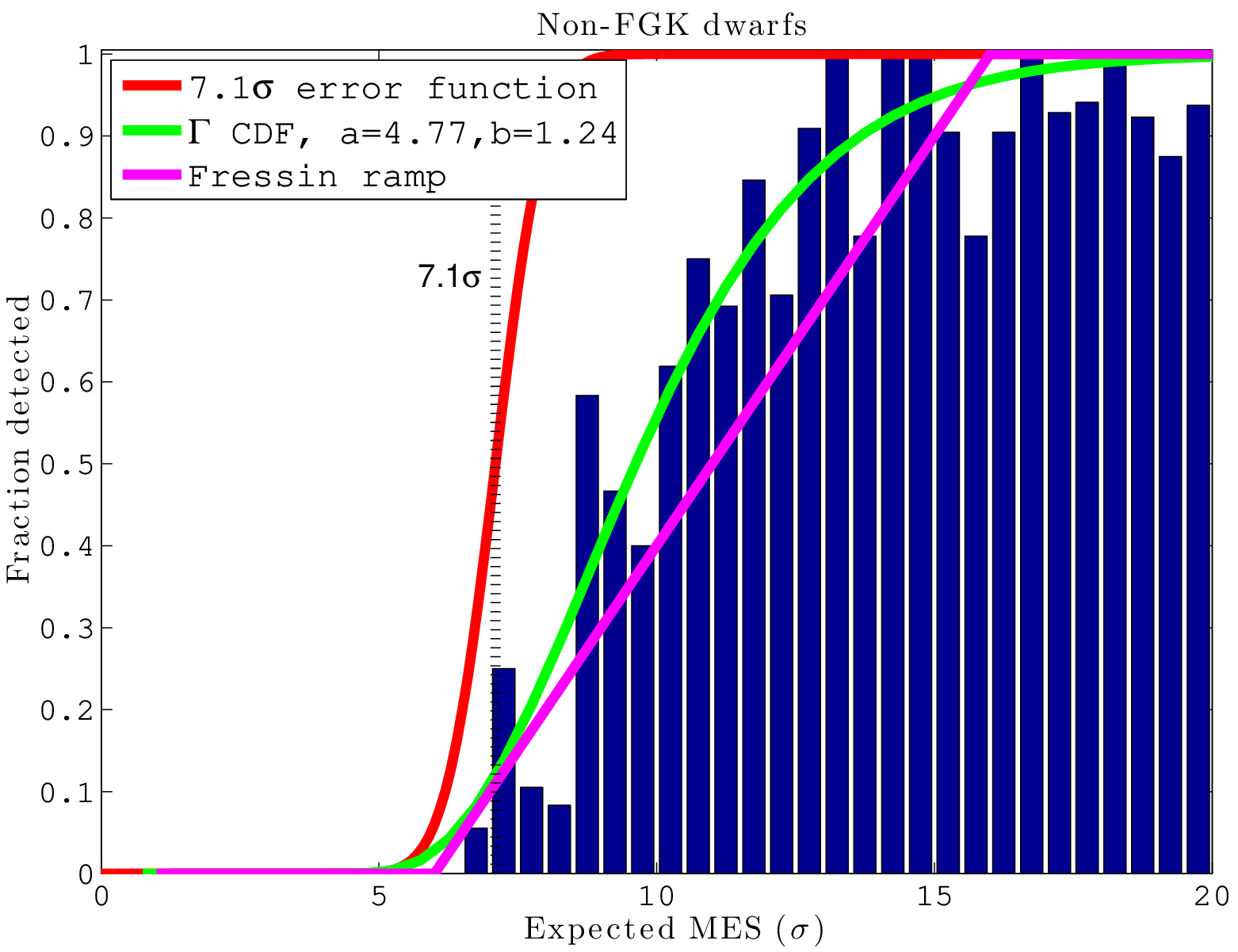}\caption{The fraction of injected signals successfully recovered by the pipeline, for the FGK dwarfs (4000K $< T_{\rm eff} <$ 7000K, log $g>4.0$; 8,579 injections in total) in the upper panel and the remaining targets (M-dwarfs and evolved stars; 2,092 injections in total) in the lower panel. In both panels, the black dotted line is the 7.1$\sigma$ pipeline threshold; the red curve is the optimum detection efficiency of the pipeline for white noise; the green curve is the $\Gamma$ function fit to the data; and the magenta line is the linear detection efficiency used by \citet{Fressin2013}. The increased noise in the lower panel is due to the significantly lower number of trial injections.}
\label{fig:sensitivitycurves}
\end{figure}

The upper panel of Figure \ref{fig:sensitivitycurves} shows the fraction of injected transit signals recovered in the FGK sample as a function of the expected MES. The detection threshold in the pipeline is a MES of 7.1$\sigma$, shown in Figure \ref{fig:sensitivitycurves} as the vertical black dotted line. The theoretical performance of the pipeline as a perfect detector dealing with broadband coloured Gaussian noise as a function of the strength of a signal's MES (in $\sigma$), is defined by the normal error function in terms of the difference between the MES and the detection threshold. That is, the detection probability of a signal is defined by the cumulative distribution function for a unit variance, zero-mean Gaussian process evaluated at the difference between the signal's MES and the detection threshold, i.e., there is a 50\% chance of detecting a transit signal with a MES at  the threshold (MES=7.1$\sigma$), an 84\% chance of detection a transit sequence 1 sigma above the threshold (MES = 8.1$\sigma$), etc. This is shown in Figure \ref{fig:sensitivitycurves} as the solid red curve. However, the fraction of injected transit signals recovered at 7.1$\sigma$ is only $\sim$25\%, and falls well below the theoretical curve for MES 6--17$\sigma$. The measured signal recoverability of the pipeline is well-characterised by a $\Gamma$ cumulative distribution function, which has the form:

\begin{equation}
	p = F(x|a,b)=\frac{1}{b^a\Gamma(a)}\int\limits_0^x t^{a-1}e^{-t/b}dt
\end{equation}

The best-fit coefficients to the sensitivity curve are $a = 4.35$, $b=1.05$, shown as the green dashed curve.  This corresponds to a 25.8\% recovery rate of 7.1$\sigma$ signals (compared to the 50\% theoretical recovery rate), and this is the sensitivity curve we use in the remainder of our analysis. In comparison, we show the detection efficiency of the remaining targets in the stellar sample (i.e., those with $T_{\rm eff}<4000$K or $>7000$K, or log $g<4.0$) in the lower panel of Figure \ref{fig:sensitivitycurves}. There are 1,762 injections, of which 1,542 were successfully recovered (the substantially lower number of trials results in the reduced smoothness of the histogram, compared to the FGK sample). As suspected, these stars have a lower detection efficiency for the same transit signal strength, which is likely due to increased masking of real signals by the presence of correlated astrophysical signals in the light curves. The best-fit coefficients to a fit of the $\Gamma$ cumulative distribution function, shown again as the green dashed curve, are $a = 4.77$, $b=1.24$, corresponding to a 12.0\% recovery rate at 7.1$\sigma$. Taking the whole stellar sample (10,341 total injections) results in best-fit coefficients of $a = 4.21$, $b=1.13$, a 23.6\% recovery rate at 7.1$\sigma$. Table \ref{tab:coeffs} summarises the results.

Briefly examining the population of signals that are not recovered by the pipeline is useful for identifying those processes in the pipeline that are implicated. Of the 1,218 injections that are not recovered, 397 have an expected MES of greater than 15. Only 95 of these have periods longer than 3 days (below which we can attribute the non-detections to the harmonic fitter described earlier) and 17 of these are recovered at an integer multiple of the injected orbital period. Of the remaining 78 signals, 46 ($\sim60$\%) show very strong stellar variability, stellar rotation or eclipsing binary signals; these light curves may not be treated well by the harmonic fitter, and/or may generate many spurious detections that cause the injected signal to be missed. Of the remaining 32 signals, 26 have fewer than six transits. Here, a significant factor contributing to the non-recovery of transit signatures is the vetoes used by TPS to discriminate between genuine transit signals and those caused by systematic noise (Seader et al. 2013, Tenenbaum et al. 2014, Seader et al. 2015). In this version of TPS, there is a documented issue that the $\chi^2$ veto is overly aggressive toward signals with a low number of transits, due to the test not taking into account the mismatch in shape between the transit signal in the data and the modelled transit template (Seader et al. 2015)\footnote{This issue will be resolved in the 9.3 version of TPS.}. Additionally, for transiting planets where the individual events have low SNR, the shape of the transit can be distorted by the noise present in the data to the extent that the transit signal can fail the vetoes. Careful tuning of the vetoes has been performed during pipeline testing to preserve as many of the `real' transit signals as possible while eliminating many tens of thousands of spurious signals, but we can see the impact of the vetoes on the signals with low numbers of transits and low SNR here. This leaves six well-behaved light curves with a reasonable number of high SNR transits for which we have no ready explanation for their non-recovery.

One comparison we can investigate is between the targets on the nominal/best focus CCD channels, and the channels at the edge of the field, which experience the worst focus. Although the focus changes across the field of view, it is relatively stable across the scientific timescales of interest, cycling on a yearly basis with the spacecraft thermal environment. Indeed, we find that there is only a slight improvement in detection efficiency for the `best' focus CCDs compared to those with the `worst' focus. Examining the FGK sample as defined above, we find best fit coefficients of $a=4.26$, $b=1.04$ (a 27.6\% recovery rate at 7.1$\sigma$) and $a=4.79$, $b=0.98$ (a recovery rate of 22.7\% at 7.1$\sigma$), respectively. This is distinct from the fact that the targets on the `worst' focus channels typically have higher CDPP values \citep{Christiansen2012} and therefore it is already more difficult to detect low SNR planetary signals in these channels; this statement is that, for a given CDPP, the detection efficiency of the pipeline is lower on the channels that are less well focussed. 

We also examined the different electronic artefacts that were sampled (e.g. variable black/bias correction, FGS crosstalk, rolling band artefacts, and start-of-line ringing; see the footnotes of Table \ref{tab:CCDs} for a description of these artefacts) and the results are delivered in Table \ref{tab:coeffs}. We find small variations in the recovery rate from the average rate for all FGK targets, with variable black/bias correction and rolling band artefact channels having a slightly higher recovery rate, and the FGS crosstalk and start-of-line ringing channels having rates comparable to the non-electronic-artefact channels examined. At this stage we are hesitant to over-interpret the slightly higher rates in some channels as meaningful, due to the relatively small number of injections and recoveries per behaviour studied, and also since the periodicity of the noise introduced by these CCDs is, like the focus changes, approximately yearly, as the targets cycle off the CCD channel with the electronic artefact and typically onto more well-behaved CCDs, before returning to the noisier channel a year later. We can expect to uncover longer period changes in sensitivity when a longer observation baseline is examined in the next experiment. 

\begin{table}[h]
\centering
\caption{$\Gamma$ function best-fit coefficients across the focal plane.}
\begin{tabular}{llllll}
\hline
\hline
Targets & Injections & Recoveries & $a$ & $b$ & R($7.1\sigma$) \\
\hline
\multicolumn{6}{c}{All channels}\\
\hline
All targets & 10341 & 9123 & 4.21 & 1.13 & 23.6\% \\
FGK targets & 8579 & 7696 & 4.35 & 1.05 & 25.8\%\\
Non-FGK targets & 1762 & 1427 & 4.77 & 1.24 & 12.0\%\\
\hline
\multicolumn{6}{c}{Subset of channels, FGK targets}\\
\hline
Nominal/best focus & 1077 & 953 & 4.26 & 1.04 & 27.6\% \\
Edge of field/worst focus & 1732 & 1553 & 4.79 & 0.98 & 22.7\% \\
Variable black/bias correction & 1350 & 1220 & 3.88 & 1.15 & 29.1\%\\
Rolling band artefacts/Moir\'{e} pattern drift & 1961 & 1750 & 3.86 & 1.12 & 28.9\%\\
FGS crosstalk & 1977 & 1791 & 4.97 & 0.91 & 23.9\%\\
Start-of-line ringing & 482 & 429 & 7.72 & 0.58 & 18.0\%\\
\hline
\hline
\end{tabular}
\label{tab:coeffs}
\end{table}

\subsection{Recovery of injected parameters}

\begin{figure}[t!]
\centering
\includegraphics[width=\columnwidth]{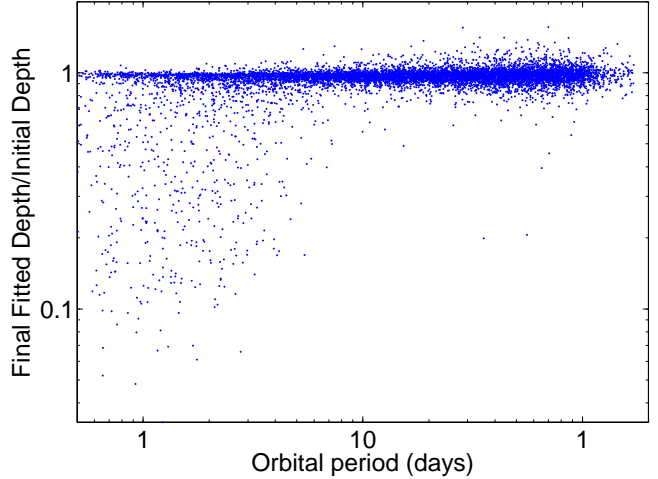}
\caption{Distribution of transit depth recovery as a function of orbital period. The majority of points at periods longer than 10 days lie around a value of 1, indicating that the pipeline is preserving the depths of the transits well. The population of points with periods shorter than 10 days shows a large scatter towards recovering shallower depths than are injected. This can be attributed largely to the action of the harmonic filter, which removes sinusoidal signals in the light curve before the data are searched for transit signals, and acts more strongly on short-period transit signals.}
\label{fig:depthvperiod}
\end{figure}

For the recovered injections, we can compare the values fitted by the pipeline to those injected. First, we examine the preservation of the injected transit depth through the pipeline, and find that the pipeline algorithms do a very good job of preserving the expected depth. We showed in Paper I that the first stage of the pipeline (generation of the aperture photometry and cotrending of the photometry to remove systematics) reduced the depths of the injected transits by a very small amount: the average measured SNR of the injected transits was 99.7\% of the injected SNR, which for individual transits (the focus of that paper) corresponds directly to a 99.7\% recovery of the injected depths. Here, we take the ratio of the final fitted transit depth of the folded transit as reported by the Data Validation (DV) pipeline module to the average measured transit depth of the injected transits. In Figure \ref{fig:depthvperiod}, we show this ratio as a function of the injected orbital period. There are two populations of points in this plot: those with periods longer than 10 days, and those with shorter periods. 

\begin{figure}[t!]
\centering
\includegraphics[width=\columnwidth]{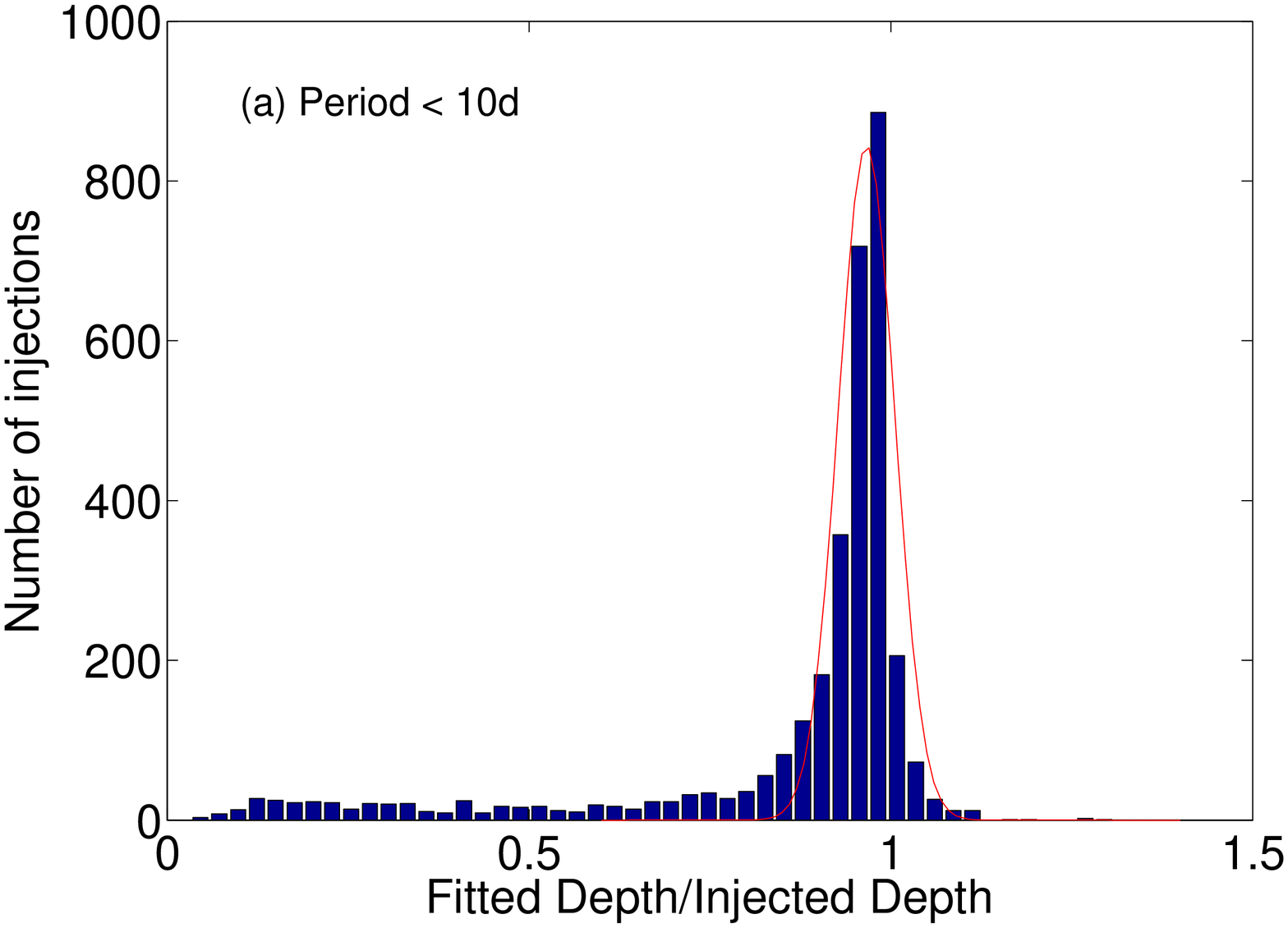}
\includegraphics[width=\columnwidth]{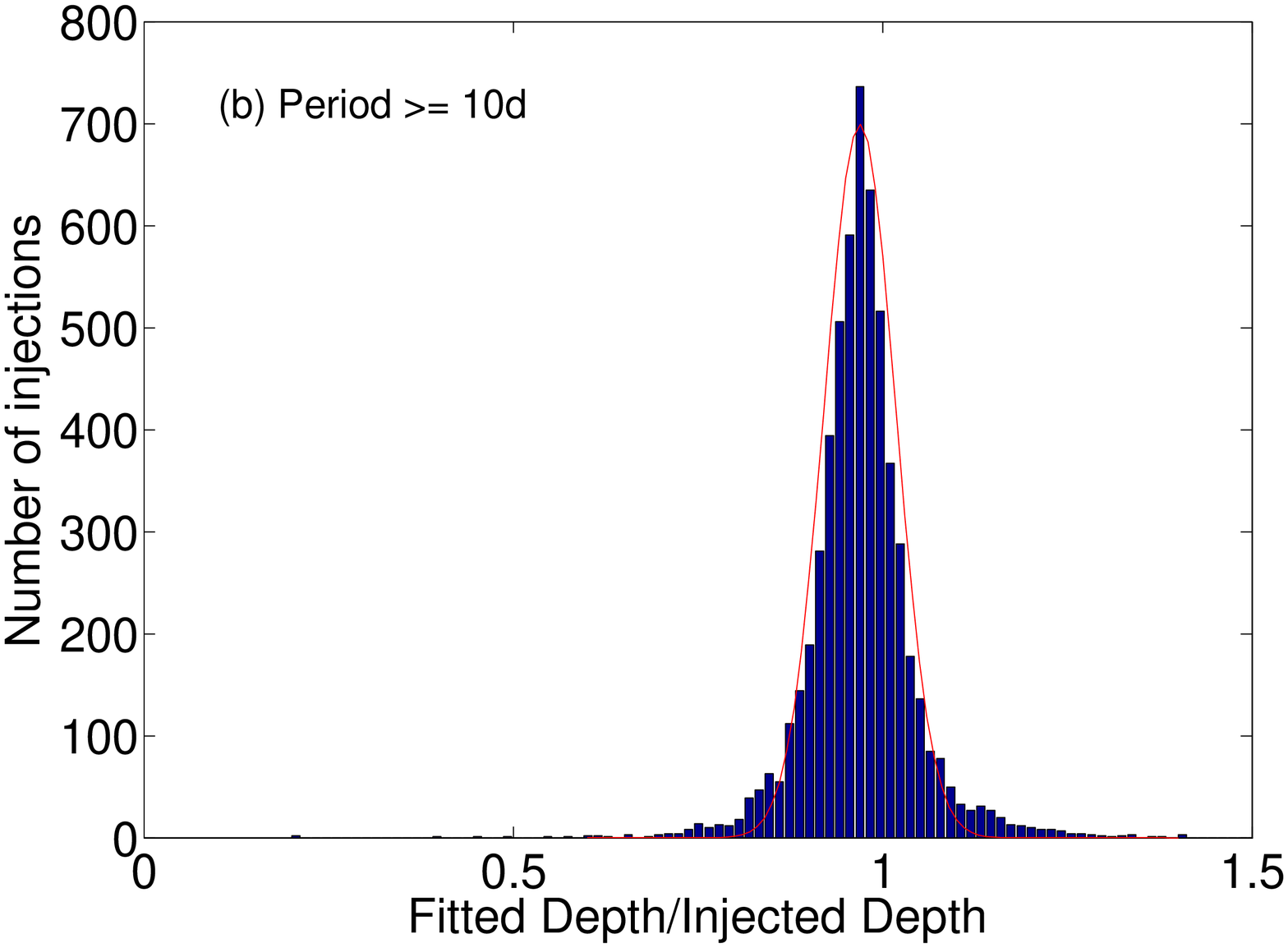}
\includegraphics[width=\columnwidth]{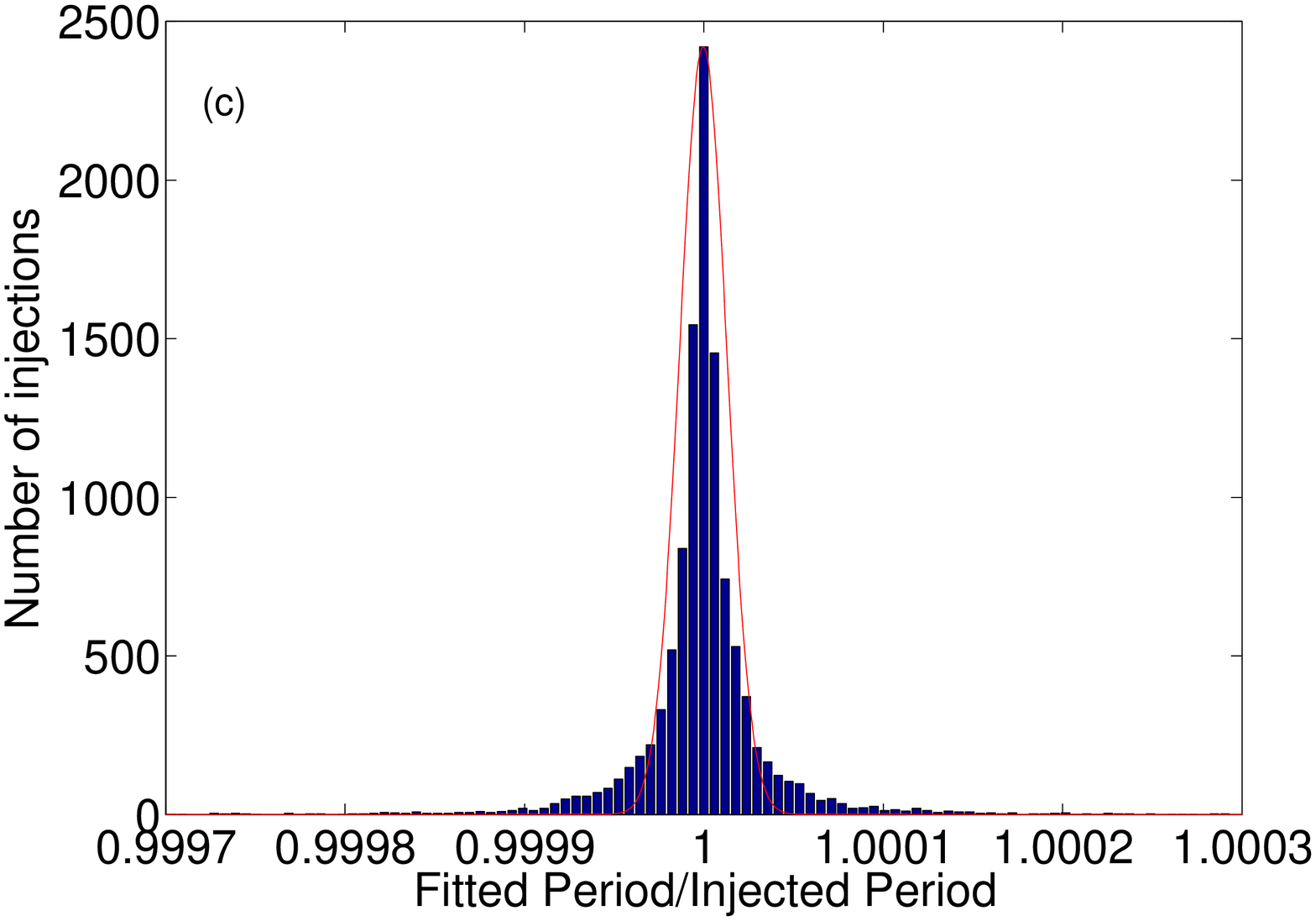}
\caption{Comparisons of the injected and fitted parameters. Panel (a) shows the ratio of the fitted depth to the injected depth for injected transit signals with periods shorter than 10 days, panel (b) for injected transit signals with periods longer than 10 days, and panel (c) shows the ratio of the fitted period to the injected period across the entire period range.}
\label{fig:comparisons}
\end{figure}

A histogram of the ratio for the shorter period population is shown in panel (a) of Figure \ref{fig:comparisons}; also shown is a best-fit normal distribution, with a median value of 97.0\% and a standard deviation of 4.0\%. However, the mean is only 87.3\%; the main reason for the reduction in measured transit depth compared to injected transit depth is the harmonic filter, as discussed earlier in this section. There, we were specifically highlighting injections which were not successfully recovered because the reduction in transit depth brought them below the detection threshold of the pipeline; here we see the population of injections that are recovered, but at a reduced depth. This has stronger implications for the inferred size of the planet population at periods shorter than 10 days which should be folded into population analyses.

The best-fit normal distribution to the longer period population is shown in panel (b) of Figure \ref{fig:comparisons}, with a mean value of 97.0\% and a standard deviation of 5.0\%. This indicates that the pipeline algorithms are reducing the depths of the longer period signals by $\sim$3\% in the latter stages, which include the harmonic filtering to remove sinusoidal signals, the whitening of the light curve to remove correlated noise, the normalising of the resulting harmonic-removed, whitened light curves, and the fitting of the transit model to the final full light curve product. This reduction results in a slight ($\sim$1.7\%) decrease in the planet radius that would be measured from the light curve than what was injected; this is considerably smaller than the typical planet radius errors, but is a systematic average decrease and should be taken into account in robust population analyses. 

Although we have validated the pipeline algorithms here in preserving the transit depth, we caution that the results should not be interpreted as comprising the total error budget on the planet radius that would be inferred from the measured transit depth. There are at least two additional source of uncertainty to consider---the stellar parameters and the dilution of the transit depth by third light in the photometric aperture. The pipeline estimates a correction for this dilution by modelling the known sources in the aperture and subtracting the appropriate flux, but this estimation is limited and constitutes an additional error on the planet radius calculation.

We also compare the measured orbital period to the injected orbital period, and find an extremely tight distribution, shown in panel (c) of Figure \ref{fig:comparisons}. Fitting a normal distribution to the data gives a mean value of 1 and a standard deviation of 1.3e-5; it is therefore very unlikely that our criteria for a `successful' recovery of matching the period to within 3\% has eliminated real detections. At the longest periods examined here ($\sim$185 days), this corresponds to an uncertainty of 3 minutes.

\section{Discussion}
\label{sec:discussion}

The most significant impact of the detection efficiency is on the calculation of the underlying planet occurrence rates. As described in Section \ref{sec:intro}, previous analyses have typically assumed some threshold or curve for the detection efficiency. \citet{Catanzarite2011} assumed 100\% completeness for signals with an SNR$>7.1\sigma$ for planets larger than $2\rm{R}_{\oplus}$ and periods less than 130 days. \citet{Borucki2011b} assumed the perfect detector described in the previous section (50\% completeness at SNR of 7.1$\sigma$), for a signal with a minimum of two transits, out to orbital periods of 138 days. \citet{Howard2012} assumed a 100\% detection efficiency for transiting signals with an SNR $>10\sigma$, for planets larger than $2\rm{R}_{\oplus}$, and periods less than 50 days, based on the decreasing rate of detection of these objects with time, and their readily apparent signals to the eye. \citet{Youdin2011} makes the same assumption, and extends the parameter space to planets down to $0.5\rm{R}_{\oplus}$. In a similar fashion, \citet{Dong2012} assume a 100\% detection efficiency for transit signals with an SNR $>8\sigma$ for planets larger than $1\rm{R}_{\oplus}$, out to periods of 250~days (and found similar results for transit signals with an SNR $>12\sigma$). \citet{Fressin2013} estimate the detection efficiency from a comparison of the distribution of their modelled false positives with the planet candidates reported in \citet{Batalha2012}, finding a linear increase in detection efficiency from 0\% at $6\sigma$ to 100\% at $16\sigma$, shown in Figure \ref{fig:sensitivitycurves} as the magenta dashed line. Since these studies have examined different planet parameter spaces, it is difficult to isolate the impact of the choice of detection efficiency on the derived occurrence rates. 

In order to examine this impact, we perform a toy-model occurrence rate analysis three times, varying only the assumption of detection efficiency in each case: for an optimistic detection efficiency assumption (the perfect detector, used by Borucki et al. 2011b\nocite{Borucki2011b}), called Detection Efficiency 1 (DE1) for the remainder of the paper; the detection efficiency curve empirically measured in this study, called DE2; and a pessimistic assumption (the linear ramp in probability of detection from 0\% at 6$\sigma$ to 100\% at 16$\sigma$ described by \citet{Fressin2013}, called DE3. All three of these detection efficiency curves are shown in Figure \ref{fig:sensitivitycurves}.

\begin{figure}[t!]
\centering
\includegraphics[width=\columnwidth]{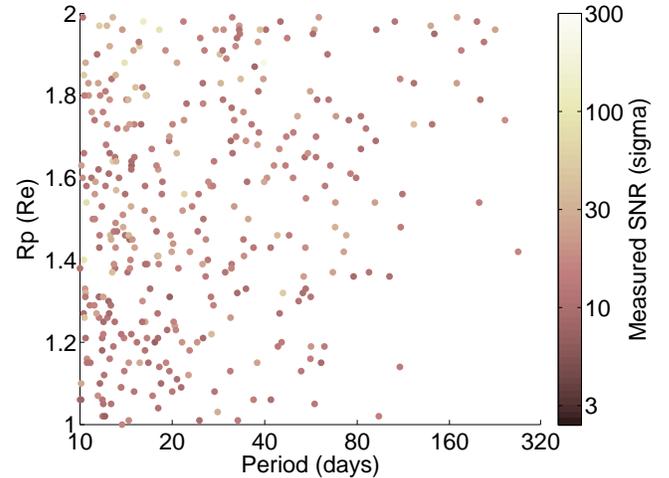}
\caption{The distribution in period and radius of the 352 planet candidates selected for this study. The colour of the point corresponds to the reported SNR at which the planet candidate was detected.}
\label{fig:planetsample}
\end{figure}

Using the (closed) Q1--Q12 Kepler Object of Interest (KOI) table at the NASA Exoplanet Archive (Rowe et al. 2015), we select all the objects classified as `Planet Candidates' with planet radii from 1--2$\rm{R}_{\oplus}$, and orbital periods from 10--320~days\footnote{http://exoplanetarchive.ipac.caltech.edu, as of December 11th, 2014}, around the FGK stellar sample as defined above. This results in 352 planet candidates, shown in Figure \ref{fig:planetsample}. There are a large number of false positives at periods $<10$~days and $>320$~days (see Fig. 8 of Tenenbaum et~al.~2013), so we exclude those candidates in this sample. We caution strongly that the Q1--Q12 KOI sample was not a uniformly-generated planet sample (see Rowe et al. 2015 for more details), and we are using it solely as a starting point to illustrate the systematic errors that arise from different assumptions in the detection efficiency, not as a starting point for deriving robust occurrence rates. 

For each of the detection efficiency assumptions, we perform the following analysis, similarly to PHM13:

\begin{enumerate}
	\item For each planet in the sample, we calculate the geometric correction, $f_g = a/R_*$, where $a$ is the semi-major axis of the orbit, and $R_*$ is the radius of the star the candidate is suspected to be transiting.
	\item Using the Q1--Q16 stellar table (the most up-to-date stellar parameters available, Huber et al. 2014) from the NASA Exoplanet Archive, we then check, for each of the $N_s$ = 152,066 FGK stars with some observations between Q1 and Q12, whether this planet would have been detected around that star. We do this by comparing the transit depth of the planet around that star $(R_p^2/R_s^2)$, where $R_p$ is the planet radius and $R_s$ is the stellar radius from the aforementioned table, to the Q1--Q12 root-mean-square (rms) of the CDPP value of that star, for the transit duration closest to that calculated for this planet\footnote{Each star has a set of Combined Differential Photometric Precision (CDPP) values, which are a measure of the noise in the target light curve as calculated by the \kepler\ pipeline in the whitened domain, for a set of timescales corresponding to transit durations of interest, 1.5--15 hours \citep{Christiansen2012}.}. We then determine the average number of transits of the planet we could expect on that star, by using the orbital period and the distribution of observations obtained for that star over the twelve quarters, and calculate the total expected SNR on average for this planet/star combination. 
	\item Applying the relevant detection efficiency curve (DE1, DE2 or DE3) we determine the probability, $p_{s}$, that this planet/star combination would have been detected by the pipeline, given that the planet transits the star.
	\item Thus for each planet, we find the fraction of stars around which it would have been detected, $f_d = (\sum_{s=1}^{N_s} p_s)/N_s$, by summing the probability for each star and dividing by the total number of stars. 
	\item Finally, we correct each planet for its incompleteness (geometric and detectability), where $C_p = f_g/f_d$ is the corrected number of planets.
\end{enumerate} 

We then calculate the occurrence rates in a similar fashion to \citet{Howard2012} and PHM13, by performing the calculation at regularly-spaced grid points in planet radius and $log$(orbital period). We divide orbital period into bins of 0.5--1.25, 1.25--2.5, 2.5--5, 5--10, 10--20, 20--40, 40--80, 80--160 and 160--320 days, and planet radius into four equal bins from 1--2$\rm{R}_{\oplus}$. In each bin, we calculate the total number of planets falling in that bin, using the corrected values described above, and divide by the total number of stars observed, $N_s$. We use the Poisson uncertainties in the original number of planets falling into each bin to determine the final uncertainties. 

\begin{figure*}[h!]  
\centering
\includegraphics[width=0.52\textwidth]{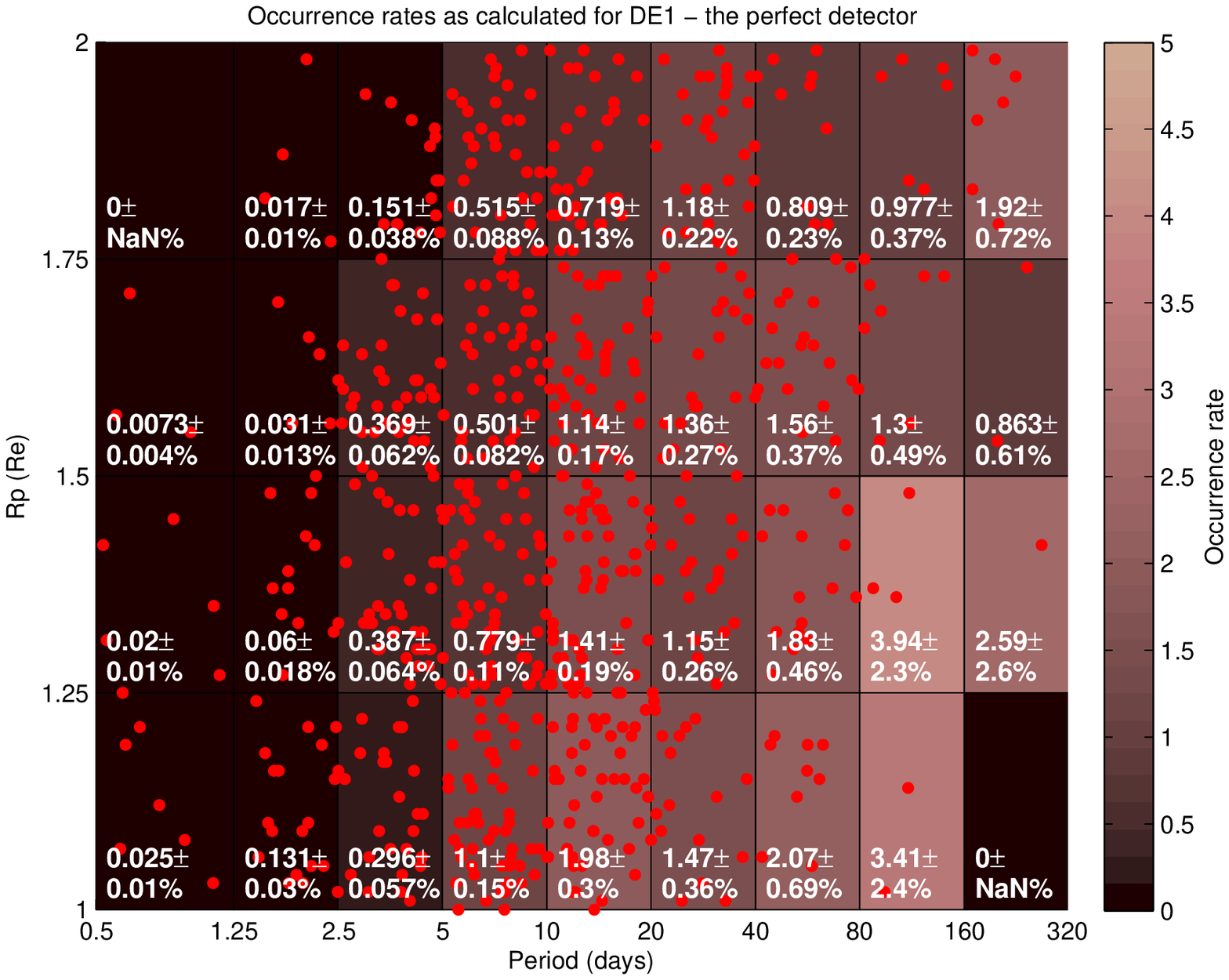}
\includegraphics[width=0.52\textwidth]{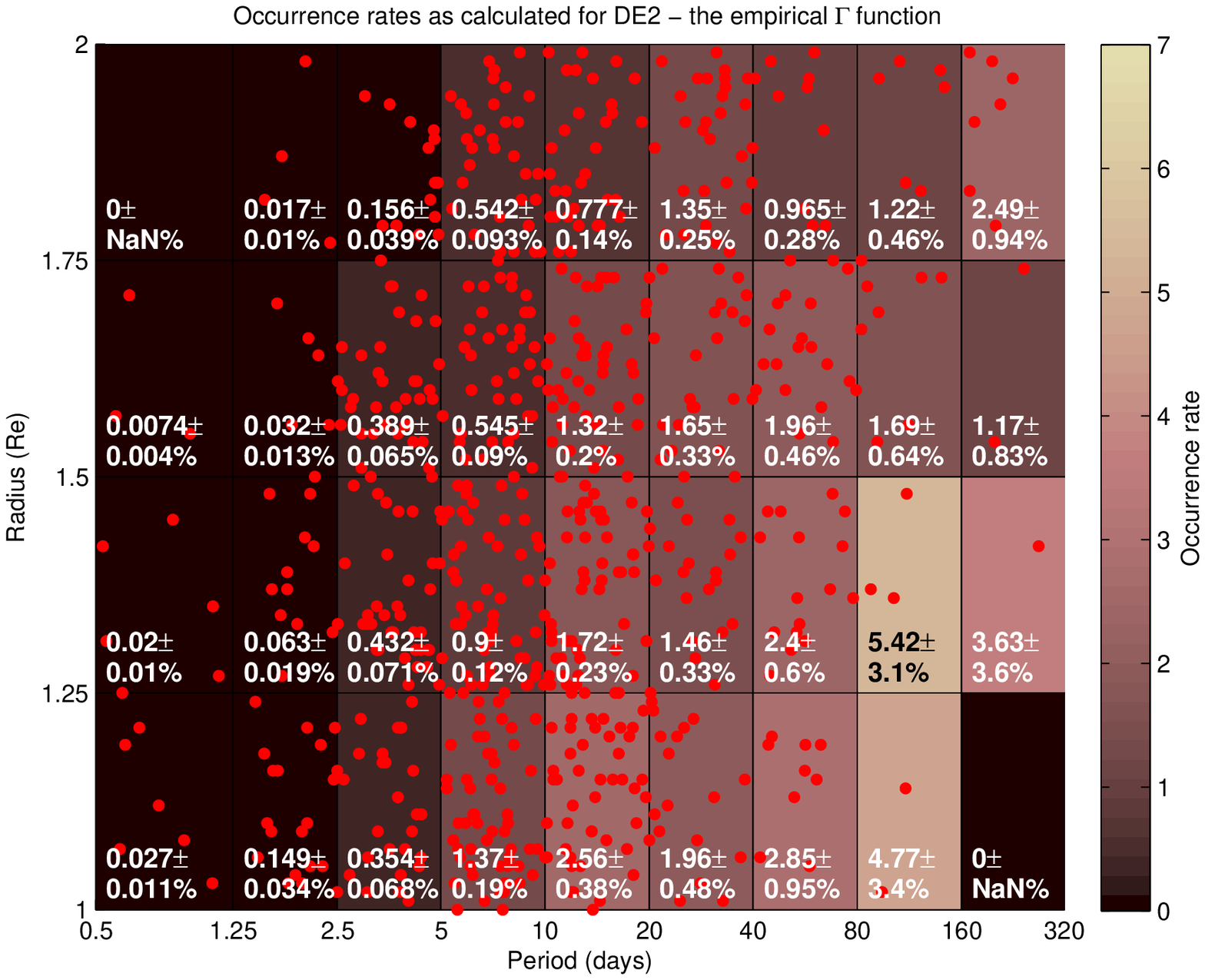}
\includegraphics[width=0.52\textwidth]{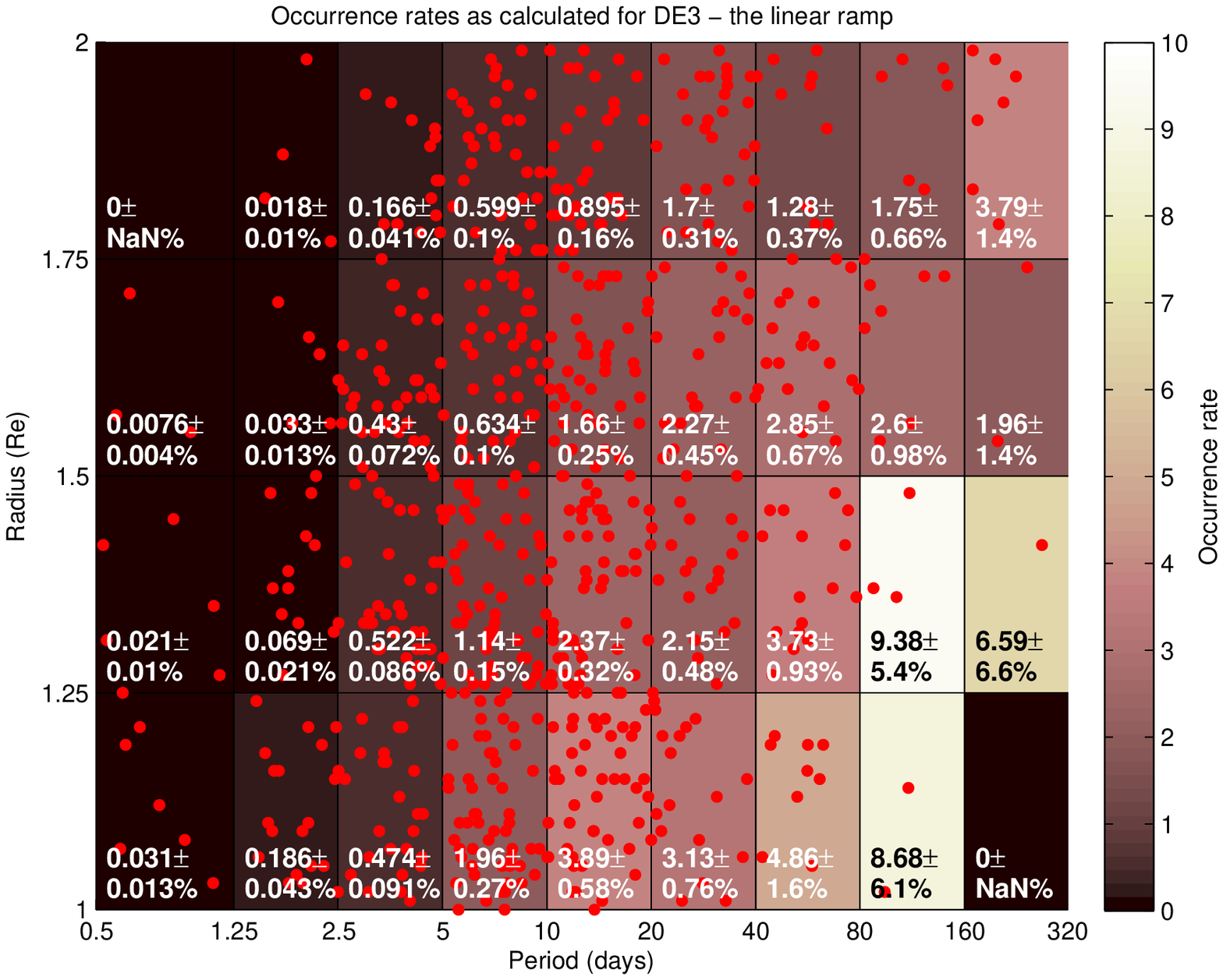}
\caption{The occurrence rates as calculated assuming the detection efficiency of the theoretical perfect detector (DE1, top), the measured efficiency (DE2, middle), and the linear ramp (DE3, bottom). The colours of the boxes scale with the occurrence rate in that box.}
\label{fig:occrate1}
\end{figure*}

The differences caused by varying the detection efficiency assumptions manifest themselves in the derived occurrence rates. Figure \ref{fig:occrate1} shows the results of the occurrence rate calculation for each of the three detection efficiency curves tested. In each case, the 701 planet candidates are plotted in red. The grid in which the occurrence rates are calculated is overlaid, and the colour of the box scales with the occurrence rate as measured in that box. The colour scale is fixed across all three images so they can be compared directly. We caution again that these are not an attempt at robustly derived occurrence rates; we have neglected many effects, such as the false positive rate of the planet sample, the uniformity of the planet sample, the reliability in the human vetting of the planet sample, and any multiplicity or systematic error in the parameters in the stellar sample. This is a simple analysis designed to examine the impact of the choice of detection efficiency curve on the derived occurrence rates, and to attempt to quantify the underlying systematic errors that result.

Qualitatively, the optimistic detection efficiency curve (DE1) assumes the pipeline detected all the planets it would have been expected to, and the derived occurrence rate is based on the number found. In contrast, the pessimistic efficiency curve (DE3) assumes the pipeline missed detections, and subsequently the number found is corrected to a higher `real' value before the occurrence rate is calculated, resulting in a significantly higher derived occurrence rate of planets. The discrepancies are most significant for orbital periods in the range 5--80 days, where the derived occurrence rates are $\ge3\sigma$ discrepant between the optimistic and pessimistic assumptions. The empirical detection efficiency curve measured in this study lies between the optimistic and pessimistic curves, as shown in Figure \ref{fig:sensitivitycurves}, and this is reflected in the derived occurrence rates, which are systematically higher than the occurrence rates derived under the optimistic case and lower than those under the pessimistic case across the entire period range.

The differences are more significant when we integrate over the radius range under consideration and derive the total occurrence rates for planets with radii between 1--2$\rm{R}_{\oplus}$. These integrated occurrence rates are given in Table \ref{tab:intrates}. For periods shorter than 5 days, most transit signals will have a very high expected MES, where the three detection efficiency curves converge, and indeed we see very little difference in the derived occurrence rates between the models. At longer periods, the occurrence rates derived with the measured detection efficiency curve are  similar to (although systematically higher than) those with the optimistic curve, only diverging at the longest periods due to the drop-off in average MES described above. However the occurrence rates derived with the pessimistic curve disagree at more than 3$\sigma$ from the optimistic or measured curves for periods 5--160 days; i.e. using the pessimistic detection efficiency curve, you would rule out the occurrence rate as derived with the measured detection efficiency curve presented here. This highlights the need for empirical measurement of the pipeline detection efficiency, and the continued re-evaluation of the pipeline performance as it is improved and applied to longer datasets; the pessimistic curve was derived from an earlier data set and earlier version of the pipeline, and it is not surprising that it is no longer applicable to the current planet population analyses.




\begin{sidewaystable}[h]
\footnotesize
\centering
\caption{Integrated occurrence rates (percentage of FGK stars with planets in this period range) across 1--2$\rm{R}_{\oplus}$, for the period ranges 0.5--1.25 days, 1.25--2.5 days, 2.5--5 days, 5--10 days, 10--20 days, 20--40 days, 40--80 days, 80--160 days and 160--320 days.}
\begin{tabular}{llllllllll}
\hline
\hline
\footnotesize{Detection Efficiency} & \footnotesize{0.5--1.25d} & \footnotesize{1.25--2.5d} & \footnotesize{2.5--5d} & \footnotesize{5--10d} & \footnotesize{10--20d} & \footnotesize{20--40d}& \footnotesize{40--80d} & \footnotesize{80--160d} & \footnotesize{160--320d}\\
\hline
DE1 (optimistic) & 0.053$\pm$0.014\% & 0.24$\pm$0.04\% & 1.20$\pm$ 0.11\% & 2.90$\pm$0.22\% & 5.26$\pm$0.40\% & 5.15$\pm$0.54\% & 6.27$\pm$0.85\% & 9.62$\pm$2.21\% & 5.37$\pm$1.69\% \\
DE2 (measured) & 0.055$\pm$0.014\% & 0.26$\pm$0.04\% & 1.33$\pm$0.12\% & 3.36$\pm$0.25\% & 6.38$\pm$0.48\% & 6.43$\pm$0.67\% & 8.17$\pm$1.10\% & 13.11$\pm$3.01\% & 7.29$\pm$2.31\% \\
DE3 (pessimistic) & 0.060$\pm$0.015\% & 0.31$\pm$0.05\% & 1.59$\pm$0.15\% & 4.33$\pm$0.32\% & 8.81$\pm$0.66\% & 9.25$\pm$0.97\% & 12.72$\pm$1.71\% & 22.42$\pm$5.15\% & 12.34$\pm$3.90\% \\
\hline
\hline
\end{tabular}
\label{tab:intrates}
\end{sidewaystable}

\section{Conclusions and further work}
\label{sec:conclusions}

We present here the first empirical measurement of the detection efficiency of the \kepler\ pipeline when detecting periodic transit signals, based on the injection and recovery of simulated transit signals injected into the calibrated pixel data. With a baseline of one year, we find that the measured detection efficiency for FGK dwarfs (4000K $< T_{\rm eff}<$ 7000K, log$g>4.0$) is well described by a $\Gamma$ function with the coefficients $a = 4.35$, $b=1.05$. However, we know from examination of the \kepler\ pipeline detections for longer baselines (see, for example, Tenenbaum et al. 2013\nocite{Tenenbaum2013}) that unexpected behaviour occurs at longer periods than those examined in this study, especially periods 300--400 days, due to the annual rotation of the targets around the \kepler\ field of view. Since one of the primary goals for the \kepler\ Mission is the measurement of the occurrence rate of planets in the habitable zone of stars like the Sun, and these longer periods encompass that parameter space, it is imperative that we extend the analysis described here to longer baselines. We plan to run a comparable transit injection experiment for the full \kepler\ observing baseline (Q1--Q17) and derive the equivalent detection efficiency curve. One particular area of study will be the detectability of multi-planet systems---the extent to which the presence of multiple periodic signals in the data and the order in which they are detected and removed by the pipeline before subsequent searches impacts the final detection efficiency

Another process that needs to be quantified is the examination of the pipeline candidates and subsequent classification as either planet candidates or false positives. For the early catalogues (up to the Q1--Q12 catalogue, Rowe et al. 2015), this classification was done entirely by a team of humans, evaluating each candidate one by one. The project is moving towards more automated methods of classification by creating algorithms to automate the decision making process (McCauliff et al. 2015, Jenkins et al. in prep). The first steps toward a completely automated process were taken in the Q16 catalogue (Mullally et~al. 2015). The final vetting process, whether human- or machine-based, introduces an additional `detection efficiency', whereby some real planet candidates may not be promoted to planet candidate status and be incorrectly classified as false positives. Our plan is to use the longer baseline run described above to also quantify the detection efficiency of the post-pipeline analysis, by reproducing the decision-making process as closely as possible, and examining the rate at which our `real' planets are discarded.

Finally, we also have the capacity to inject the simulated transit signal at a location offset from the target star, allowing us to simulate false positive signals (i.e. due to eclipsing binaries along the line of sight). By allocating some number of targets in the longer baseline experiment to study this, we will be able to examine the ability of the pipeline centroid analysis to identify and discard these false positives, and to identify the parameter space in which this identification is reliable.


\acknowledgments

Funding for the \kepler\ Discovery Mission is provided by NASA's Science Mission Directorate. The authors acknowledge the efforts of the Kepler Mission team for obtaining the calibrated pixels, light curves and data validation diagnostics data used in this publication. These data products were generated by the Kepler Mission science pipeline through the efforts of the Kepler Science Operations Center and Science Office. The Kepler Mission is lead by the project office at NASA Ames Research Center. Ball Aerospace built the Kepler photometer and spacecraft which is operated by the mission operations center at LASP. These data products are archived at the Mikulski Archive for Space Telescopes and the NASA Exoplanet Archive. JLC is supported by NASA under award No. GRNASM99G000001.

\appendix




\clearpage
\end{document}